\documentclass[prd,tightenlines,nofootinbib,twocolumn]{revtex4}
\usepackage{epsfig,color,eso-pic}
\usepackage{amsmath,amscd,amssymb}
\usepackage{graphics,bigints}
\usepackage{diagbox}

\newcommand{\imu}{{\rm i}}

\newcommand{\Vek}[1]{{\boldsymbol#1}}

\newcommand{\zr}[1]{\mbox{\hspace*{#1em}}}
\newcommand{\ID}{\mbox{{\sf 1}\zr{-0.16}\rule{0.04em}{1.55ex}\zr{0.1}}}

\usepackage[normalem]{ulem}

\begin{document}

\title{Winding number dependence of quantum vortex energies at one-loop}

\author{N. Graham$^{a)}$, H. Weigel$^{b)}$}

\affiliation{
$^{a)}$Department of Physics, Middlebury College
Middlebury, VT 05753, USA\\
$^{b)}$Institute for Theoretical Physics, Physics Department,
Stellenbosch University, Matieland 7602, South Africa}

\begin{abstract}
We compute the one-loop vacuum polarization energies of
Abrikosov-Nielsen-Olesen (ANO) vortices with topological charge $n$
in scalar electrodynamics, for the BPS case of equal gauge and scalar
masses.  This calculation allows us to investigate the relationship
between the winding number and the quantum-corrected vortex energy,
which in turn determines the stability of higher winding
configurations against decay into configurations with unit winding.
While the classical energy is proportional to $n$, we find that 
the vacuum polarization energy is negative and approximately
proportional to $n-1$ with a small constant offset.
\end{abstract}

\maketitle

\section{Introduction}

In scalar electrodynamics with spontaneous symmetry breaking, the
Abrikosov-Nielsen-Olesen (ANO) vortex
\cite{Abrikosov:1956sx,ABRIKOSOV1957199,Nielsen:1973cs} 
is an axially symmetric configuration with topological soliton
structure in the transverse plane.
In the regular gauge, the fields vanish at the center and the winding number~$n$ 
counts the mapping of the phase of the complex scalar field at spatial
infinity onto  the unit circle.  This winding number, which we take to
be positive throughout, corresponds to a quantized magnetic flux
running along the string.

From the perspective of particle physics, we can view this solution as
the analog in two space dimensions of a 't Hooft-Polyakov 
magnetic monopole \cite{tHooft:1974kcl,Polyakov:1974ek}. So-called
cosmic string solutions can also emerge from a U(1) subgroup of the SU(2) weak interactions 
\cite{Vachaspati:1992fi,Achucarro:1999it,Nambu:1977ag,Hindmarsh:1994re,Kibble:1976sj,Shellard:1994}.  
Here the scalar field is the scalar Higgs doublet while the gauge fields are the massive vector 
bosons $W^{\pm}$ and $Z^0$. In condensed matter physics, these vortices represent the 
penetration of magnetic flux through a superconductor with the scalar field being
the condensate order parameter. The ANO model is characterized by the ratio of
the charged scalar mass to the effective mass acquired by the
gauge field through spontaneous symmetry breaking.  The inverses of
these masses correspond to the superconducting coherence length and the
London penetration depth respectively, where the former reflects the
attractive self-interaction of the condensate of Cooper pairs, while
the latter corresponds to the exponential Meissner suppression of
electromagnetic fields in the condensate, leading to superconductivity
\cite{Tinkham:1996}.

The Bogomolny-Prasad-Sommerfeld (BPS) \cite{Bogomolny:1975de,Prasad:1975kr}
case, in which the masses of the gauge and scalar fields are equal, is of particular 
interest because the classical mass is proportional to the winding
number. As a result, the quantum correction to the classical energy,
however small, determines whether higher winding number configurations
are energetically stable against decay into vortices with unit winding. 
Our investigation provides the first nontrivial calculation of vacuum 
polarization energy (VPE) quantum corrections for a topological soliton 
with varying winding number in four spacetime dimensions.

In the scattering theory formalism, topological effects are captured in
the behavior of the phase shift in the presence of zero-mode and
threshold bound  states, as expressed by Levinson's theorem and its
generalizations to situations with topological boundary conditions
\cite{Niemi,Boyanovsky,super1d,qnum}.  Since Levinson's theorem
compares the phase shift at threshold to the phase shift at infinite
wave number, it captures the effects of the soliton's global topology at 
short distance. For gauge theory solitons in three space 
dimensions, such as the ANO vortex and 't Hooft-Polyakov monopole 
one encounters a generalization of this behavior; making the gauge 
field go to zero at large distances, as is required for the scattering 
theory assumption of asymptotically free interactions, necessarily 
introduces singularities in the gauge field at the origin. These 
singularities disappear in gauge-invariant quantities, however, so 
that the soliton has finite energy density everywhere.

These singularities require the subtraction of quantities that
formally vanish in the scattering theory calculation, but in practice
cancel quadratically divergent contributions in intermediate results
\cite{Graham:2019fzo}.  This subtraction enforces the cancellation of
the superficial quadratic divergence of the gauge vacuum polarization,
leaving only the logarithmic divergence, which is then accounted for
through standard renormalization.  This subtlety was first addressed in
Ref.~\cite{Baacke:2008sq} using an ad-hoc subtraction, while
the more recent work in Ref.~\cite{Graham:2019fzo}
provides a simpler and more systematic algorithm
for ensuring gauge invariance, which also leads to more reliable
numerical results.  A key test of the validity of this
approach is that it renders the result insensitive to the choice
of a minimal radius in the scattering calculation; one cannot
numerically extend all the way to the origin because the gauge fields
diverge there, but the subtracted quantities remain well-behaved in
that limit, reflecting the absence of singularities in measurable
quantities.

In the case of string solutions like the ANO vortex 
these subtleties can be further obfuscated by the slow convergence of the
sum over partial waves.  As a result, with insufficient numerical
computation unrenormalized quantities can appear finite
\cite{Pasipoularides:2000gg}, meaning that the effects of
renormalization will make them appear to diverge, when the true
calculation shows the opposite behavior.  In this work, we use the
technical formulation established in Ref.~\cite{Graham:2019fzo}
combined with the ``fake boson'' formalism
\cite{Farhi:2001kh} to precisely subtract Born approximations to the
scattering data that can then be added back in as renormalized Feynman
diagrams implementing a modified on-shell renormalization scheme.  

\section{ANO Vortices}

The classical vortices are constructed from the 
Lagrangian of scalar electrodynamics
\begin{equation}
\mathcal{L}=-\frac{1}{4}F_{\mu\nu}F^{\mu\nu}+|D_\mu\Phi|^2
-\frac{\lambda}{4}\left(|\Phi|^2-v^2\right)^2\,,
\label{eq:lag1}\end{equation}
where $F_{\mu\nu}=\partial_\mu A_\mu-\partial_\nu A_\mu$
and $D_\mu\Phi=\left(\partial_\mu-\imu e A_\mu\right)\Phi$.

\begin{figure}[ht]
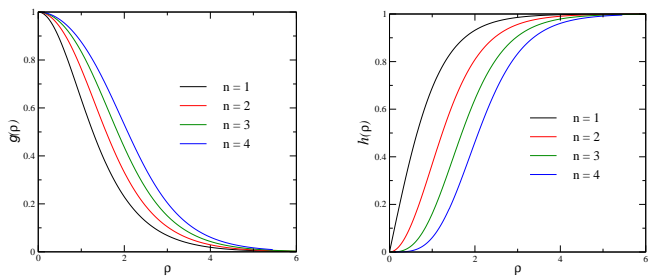

\centerline{
\epsfig{file=hhh.eps,width=0.45\linewidth}\hfill
\epsfig{file=ggg.eps,width=0.45\linewidth}}
\caption{\label{fig:class}
(color online) Classical profile functions of the vortex 
profile functions for the BPS case.}
\end{figure}

The vortex profiles in the singular gauge are
\begin{equation}
\Phi_S=vh(\rho) 
\qquad {\rm and}\qquad 
\Vek{A}_S=nv\hat{\Vek{\varphi}}\,\frac{g(\rho)}{\rho}
\label{eq:string1}\end{equation}
where $\rho=evr$ is dimensionless while $r$ is the physical coordinate. The 
winding number $n$ is the essential topological quantity. In the BPS case with
$\lambda=2e^2$, the energy functional is minimized when the profile functions
obey the first-order differential equations
\begin{equation}
g^\prime=\frac{\rho}{n}(h^2-1)
\qquad {\rm and}\qquad
h^\prime=\frac{n}{\rho}gh\,.
\label{eq:cldeqBPS}\end{equation}
and the boundary conditions
\begin{equation}
h(0)= 1 - g(0)=0 \hbox{~and~}
\lim_{\rho\to\infty}h(\rho) =
1-\lim_{\rho\to\infty}g(\rho)=1\,.
\label{eq:bc}\end{equation}
Numerical solutions are displayed in Fig.\ref{fig:class}.
The resulting energy is then proportional to the winding number,
$E_{\rm cl}=2\pi n v^2$.

\bigskip

\section{Quantum Theory}

To quantize the theory we introduce fluctuations about the vortex via
\begin{equation}
\Phi=\Phi_S+\eta \qquad {\rm and}\qquad 
A^\mu=A_S^\mu+a^\mu
\label{eq:fluct}\end{equation}
and extract the harmonic terms in the fluctuations $\eta$ and $a^\mu$.
The gauge is fixed by adding an $R_\xi$ type Lagrangian that 
cancels the $\eta\partial_\mu a^\mu$ and $\eta^\ast\partial_\mu a^\mu$ terms,
\begin{equation}
\mathcal{L}_{\rm gf}=-\frac{1}{2}\left[\partial_\mu a^\mu
+\imu e\left(\Phi_S\eta^\ast-\Phi_S^\ast\eta\right)\right]^2\,.
\label{eq:laggf}\end{equation}
We still have to account for ghost contribution to the VPE associated
with this gauge fixing,
$\mathcal{L}_{\rm gf}=-\frac{1}{2}G^2$.
The infinitesimal gauge transformations reads
\begin{equation}
A^\mu\to A^\mu+\partial \chi\,, \qquad 
\Phi_S+\eta\to \Phi_S+\eta+\imu e\chi(\Phi_S+\eta)
\end{equation}
so that
$\eta\to\eta +\imu e\chi(\Phi_S+\eta)\,.$
Then
\begin{equation}
\frac{\partial G}{\partial \chi}\Big|_{\chi=0}
=\partial_\mu\partial^\mu+e^2\left(2|\Phi_S|^2+\Phi_S\eta^\ast+\Phi_S^\ast\eta\right)\,.
\label{eq:fp1}\end{equation}
This induces the ghost Lagrangian \cite{Lee:1994pm,Rebhan:2004vu}
\begin{equation}
\mathcal{L}_{\rm gh}=\overline{c}\left(\partial_\mu\partial^\mu+2e^2|\Phi_S|^2\right)c +
\mbox{non-harmonic terms}\,.
\label{eq:fp2}\end{equation}
The corresponding VPE is that of a Klein Gordon field with mass $\sqrt{2} ev$ 
in the background potential $2v^2(h^2-1)$, which must be subtracted with a factor of 
negative two from the VPE obtained for the gauge and scalar fields.
Since $D_0\Phi_S=0$ and $D_3\Phi_S=0$, the temporal and longitudinal
components $a_0$ and $a_3$ fully decouple, contributing 
$$
-\frac{1}{2}\left[\partial_\mu a_0\partial^\mu a^0+\partial_\mu a_3\partial^\mu a^3\right]
+|\Phi_S|^2\left[a_0a^0+a_3a^3\right]
$$
to the Lagrangian. These fluctuations are subject solely to the background
potential $2(|\Phi_S|^2-v^2)$, which is exactly the same as
that of the ghosts.  As a result, the non-transverse and ghost contributions to the VPE cancel each other. Of course,
this just reflects the fact that the free electromagnetic field only has two degrees of freedom.

After canceling the non-transverse gauge fluctuations against the ghost contribution, we 
end up with the truncated Lagrangian 
\begin{align}
\mathcal{L}^{(2)}
&=\frac{1}{2}\sum_{n=1,2}\left(\partial_\mu a_n\right)\left(\partial^\mu a_n\right)
-e^2|\Phi_S|^2\sum_{n=1,2}a_n^2\cr
&\hspace{0.5cm}
+|\dot{\eta}|^2-|\partial_3\eta|^2+\sum_{n=1,2}(D_n\eta)^\ast(D^n\eta) \cr
&\hspace{0.5cm}
-e^2\left[3|\Phi_S|^2-v^2\right]|\eta|^2\cr
&\hspace{0.5cm}
+2\imu e\sum_{n=1,2}a_n\left[\eta^\ast \left(D^n\Phi_S\right)-\eta \left(D^n\Phi_S\right)^\ast\right]\,.
\label{eq:fluc}\end{align}
Essentially we have simplified the quantum gauge theory to that of four real scalar fields:
$a_1$, $a_2$, ${\sf Re}(\eta)$ and ${\sf Im}(\eta)$.

\section{Vacuum polarization energy}

To compute the VPE we will employ spectral
methods \cite{Graham:2009zz} based on the scattering theory for quantum fluctuations about the potential 
induced by the vortex. To formulate the scattering problem, we employ
a partial wave decomposition using
the complex combinations
\begin{equation}
a_{+}=\sqrt{2}\imu{\rm e}^{-\imu\omega t}\sum_\ell a_\ell(\rho){\rm e}^{\imu \ell \varphi}
\hbox{~and~}
\eta={\rm e}^{-\imu\omega t}\sum_\ell\eta_\ell(\rho){\rm e}^{\imu \ell \varphi}
\label{eq:partialwave}
\end{equation}
and similarly for $a_{-}$ and $\eta^\ast$, leading to a $4\times4$ scattering problem for the radial 
functions. For profile functions obeying Eq.~(\ref{eq:cldeqBPS}),
this problem decouples into two $2\times2$ systems,
with the one for $a_{-}$ and $\eta^\ast$
being the same as that of $a_{+}$ and $\eta$. Hence is suffices to compute the VPE of the
latter and double it. 
The scattering problem is set up in terms of the Jost solution
$\mathcal{F}_\ell$ by introducing
\begin{equation}
\begin{pmatrix}\eta^{(1)}_\ell & \eta^{(2)}_\ell \cr 
a^{(1)}_{\ell+1} & a^{(2)}_{\ell+1} \end{pmatrix}=\mathcal{F}_\ell\cdot\mathcal{H}_\ell\,,
\mathcal{H}_\ell=\begin{pmatrix}H^{(1)}_{\ell}(q\rho)& 0
\cr 0& H^{(1)}_{\ell+1}(q\rho)\end{pmatrix}\,,
\label{eq:param}\end{equation}
where the superscripts on the left-hand side refer to the two possible scattering 
channels when imposing the boundary condition $\lim_{\rho\to\infty}\mathcal{F}_\ell=\ID$.
The Hankel functions $H^{(1)}$ parameterize outgoing cylindrical waves.
In matrix form, the scattering differential equations for
(dimensionless) imaginary momentum
$t=\imu \sqrt{\omega^2-2e^2v^2}/(ev)^2$ read
\begin{equation}
\frac{\partial^2}{\partial\rho^2}\mathcal{F}_\ell
=-\frac{\partial}{\partial\rho}\mathcal{F}_\ell
-2\left(\frac{\partial}{\partial\rho}\mathcal{F}_\ell\right)\cdot\mathcal{Z}_\ell
+\frac{1}{\rho^2}\left[\mathcal{L}_\ell,\mathcal{F}_\ell\right]
+\mathcal{V}_\ell\cdot\mathcal{F}_\ell\,.
\label{eq:jostdeq}\end{equation}
The angular momenta enter via the logarithmic derivative matrix for the analytically 
continued Hankel functions
\begin{equation}
\mathcal{Z}_\ell=\begin{pmatrix}
\frac{|l|}{\rho}-t\,\frac{K_{|l|+1}(t\rho)}{K_{|l|}(t\rho)} & 0 \cr
0 & \frac{|l+1|}{\rho}-t\,\frac{K_{|l+1|+1}(t\rho)}{K_{|l+1|}(t\rho)}
\end{pmatrix}
\end{equation}
and
\begin{equation}
\mathcal{L}_\ell=\begin{pmatrix}\ell^2 & 0 \cr 
0 & (\ell+1)^2 \end{pmatrix}\,,
\label{eq:angmom}\end{equation}
and the potential matrix is
\begin{equation}
\mathcal{V}_\ell=\begin{pmatrix}
3(h^2(\rho)-1)+\frac{n^2g^2(\rho)-2n\ell g(\rho)}{\rho^2}
& \sqrt{2}d(\rho)\cr
\sqrt{2}d(\rho)& 2(h^2(\rho)-1) \end{pmatrix}\,.
\label{eq:potmat}\end{equation}
We then use Eq.~(\ref{eq:jostdeq}) to compute the Jost function, which
is given by $\nu_\ell(t)=\lim_{\rho\to0}{\rm ln}
\,{\rm det}\left[\mathcal{F}_\ell\right]$.

The standard procedure to determine the Born approximations, which are
needed to regularize the ultraviolet divergences, fails when $g(0)\ne0$. 
To perform the Born subtractions without the singular terms, we introduce
\begin{equation}
\overline{\mathcal{V}}=\begin{pmatrix}
3(h^2(\rho)-1) & \sqrt{2}d(\rho)\cr
\sqrt{2}d(\rho)& 2(h^2(\rho)-1) \end{pmatrix}
\label{eq:potmat1}\end{equation}
and iterate 
\begin{equation}
\frac{\partial^2}{\partial\rho^2}\overline{\mathcal{F}}_\ell=-\frac{\partial}{\partial\rho}\overline{\mathcal{F}}_\ell
-2\left(\frac{\partial}{\partial\rho}\overline{\mathcal{F}}_\ell\right)\cdot\mathcal{Z}_\ell
+\frac{1}{\rho^2}\left[\mathcal{L}_\ell,\overline{\mathcal{F}}_\ell\right]
+\overline{\mathcal{V}}\cdot\overline{\mathcal{F}}_\ell
\label{eq:barjostdeq}\end{equation}
according to the expansion
$\overline{\mathcal{F}}_\ell=\ID+\overline{\mathcal{F}}_\ell^{(1)}+\overline{\mathcal{F}}_\ell^{(2)}+\ldots$,
where the 
superscript refers to the order of $\overline{\mathcal{V}}$. From the derivation in Ref.~\cite{Graham:2019fzo}
we expect that
\begin{align}
\left[\nu(t)\right]_V&=
\lim_{\genfrac{}{}{0pt}{}{L\to\infty}{\rho_{\rm min}\to0}}\left\{
\sum_{\ell=-L}^{L}\left[\nu_\ell(t)-\overline{\nu}^{(1)}_\ell(t)
-\overline{\nu}^{(2)}_\ell(t)\right]_{\rho_{\rm min}} 
\right. \nonumber \\ \label{eq:sing1} &\left.\qquad
-n^2\int_{\rho_{\rm min}}^\infty \frac{d\rho}{\rho}\,g^2(\rho)\right\}
\end{align}
approaches $\displaystyle
\frac{n^2}{12t^2}\int_0^\infty
\frac{d\rho}{\rho}\,\left(\frac{dg(\rho)}{d\rho}\right)^2$
as $t\to \infty$.

The subtraction of the integral in Eq.~(\ref{eq:sing1})
cancels the superficial quadratic divergence in the VPE. 
By this subtraction we restore gauge invariance. The full computation of the 
VPE requires us to multiply the subtracted Jost function by $t$, capturing the effect of the
translation-invariant direction, and integrate \cite{Graham:2001dy}. 
The right-hand side of Eq.~(\ref{eq:sing1}) then leads to the
logarithmic divergence associated with gauge field renormalization. This divergence
is most conveniently treated within the fake boson formalism \cite{Farhi:2001kh} which
takes advantage of the fact that the second-order Born term for a scalar field also 
induces a logarithmic divergence. To be precise, we consider scattering of a boson about 
the potential $V_f=3e^2v^2(\tanh^2(\kappa evr)-1)$, for which
$\overline{\nu}^{(2)}_\ell(t)$ is the second-order contribution 
to the Jost function on the imaginary momentum axis. We take $\kappa$
as a free parameter to later test our numerical simulation, since the final 
result for VPE should not depend on a particular choice.
This subtraction is calibrated by defining
\begin{eqnarray}
c_B &=& -\frac{e^2}{6}\frac{\int_0^\infty rdr\, F_{\mu\nu}F^{\mu\nu}}
{\int_0^\infty r dr V_f^2} \cr
&=& -\frac{n^2}{3}\frac{\int_0^\infty \rho d\rho
\left(\frac{g^\prime(\rho)}{\rho}\right)^2}
{\int_0^\infty \rho d\rho\, \left[3(\tanh^2(\kappa \rho)-1)\right]^2}
\label{eq:CB}\end{eqnarray}
so that the scattering contribution to the VPE is
\begin{equation}
E_{\rm VPE}^{\rm scat.}=\frac{1}{2\pi}\int_{\sqrt{2}}^\infty tdt\,
\left[\left[\nu(t)\right]_V-c_B \overline{\nu}^{(2)}(t)\right]\,.
\label{eq:Escat}
\end{equation}

To identify the subtraction in Eq.~(\ref{eq:Escat}) in terms of
Feynman diagrams
\cite{Becchi:1974md,Irges:2017ztc} we consider the Lagrangian
with four real fields $\phi^{\rm t}=(\eta_1,\eta_2,a_1,a_2)$
\begin{equation}
\mathcal{L}=\frac{1}{2}\left(\partial_\mu \phi^{\rm t}\right)\left(\partial^\mu \phi\right)
-\frac{1}{2} \phi^{\rm t} M^2 \phi - \phi^{\rm t} V \phi \,,
\label{eq:LagFD}
\end{equation}
where $M=\sqrt{2}ev$ is the mass of both the gauge and scalar
field fluctuations. The Cartesian components of the 
gauge fields have been defined above, and
$\eta=\left(\eta_1+\imu\eta_2\right)/\sqrt{2}$.
The potential matrix is given by $V=V_0+V_1+V_2$ with
\onecolumngrid
\begin{equation}
V_0=\begin{pmatrix}\frac{3\lambda}{4}\left(\Phi^2_S-v^2\right) &
0 & \sqrt{2}e^2\,\hat{\Vek{x}}\cdot\Vek{A}_S\Phi_S & 
\sqrt{2}e^2\,\hat{\Vek{y}}\cdot\Vek{A}_S\Phi_S\\[1mm]
0 & \frac{3\lambda}{4}\left(\Phi^2_S-v^2\right) &
-\sqrt{2}e\,\hat{\Vek{x}}\cdot\Vek{\nabla} \Phi_S &
-\sqrt{2}e\,\hat{\Vek{y}}\cdot\Vek{\nabla} \Phi_S\\[1mm]
\sqrt{2}e^2\,\hat{\Vek{x}}\cdot\Vek{A}_S\Phi_S & 
-\sqrt{2}e\,\hat{\Vek{x}}\cdot\Vek{\nabla} \Phi_S & 
e^2\left(\Phi^2_S-v^2\right) & 0  \\[1mm]
\sqrt{2}e^2\,\hat{\Vek{y}}\cdot\Vek{A}_S\Phi_S  & 
-\sqrt{2}e\,\hat{\Vek{y}}\cdot\Vek{\nabla}\Phi_S &
0 & e^2\left(\Phi^2_S-v^2\right)
\end{pmatrix} \,,
\label{eq:PotMatrix0}
\end{equation} \twocolumngrid
\noindent and
\begin{equation}
V_1=e\begin{pmatrix} 0 & 1 & 0 & 0 \\[0mm] -1 & 0 & 0 & 0 \\[0mm]
0 & 0 & 0 & 0 \\[0mm] 0 & 0 & 0 & 0\end{pmatrix}\, \Vek{A}_S\cdot\Vek{\nabla}
\end{equation}
and
\begin{equation}
V_2=\frac{e^2}{2}\begin{pmatrix} 1 & 0 & 0 & 0 \\[0mm] 0 & 1 & 0 & 0 \\[0mm]
0 & 0 & 0 & 0 \\[0mm] 0 & 0 & 0 & 0\end{pmatrix}\, \Vek{A}_S\cdot\Vek{A}_S
\label{eq:PotMatrix1}
\end{equation}
Here we have separated out $V_1$ and $V_2$ since they relate to the singular terms in the
scattering problem, while $V_0$ is the $4\times4$ representation of
$\overline{\mathcal{V}}$.  The renormalization program via Feynman
diagrams in dimensional regularization is carried out with the full
potential matrix $V$, while the subtractions should only involve $V_0$
supplemented by the wave-function renormalization of the gauge boson,
which in turn is simplified by the fake boson trick.

To form the diagrammatic expansion,
we Taylor expand the effective action $\frac{\imu}{2}{\rm Tr}\left[\partial^2+M^2+2V\right]$
for the four real scalar fields. The contribution linear in $V_0$ is the scalar tadpole and corresponds 
to the subtraction of $\overline{\nu}_\ell^{(1)}$ in Eq.~(\ref{eq:sing1}). This tadpole diagram 
is fully canceled by a counterterm proportional to $\int d^4x\, \left(|\Phi|^2-v^2\right)$.
The terms quadratic in $V_0$ correspond to $\overline{\nu}_\ell^{(2)}$ in Eq.~(\ref{eq:sing1}) and
are logarithmically divergent. These divergences are canceled by counterterms proportional to 
$\int d^4x\, \left(|\Phi|^2-v^2\right)^2$ and $\int d^4x\,|D_\mu\Phi|^2$. The contributions linear in 
$V_2$ and quadratic in $V_1$ combine such that the individual quadratic divergences cancel and
the subleading logarithmic divergence matches that of the right hand side of Eq.~(\ref{eq:sing1}).
The logarithmic divergences from 
$V_0\otimes V_2$ and $V_0\otimes V_1\otimes V_1$ cancel, as do those from 
$V_2\otimes V_2$, $V_2\otimes V_1\otimes V_1$ and $V_1\otimes
V_1\otimes V_1\otimes V_1$, so the subtractions discussed after Eq.~(\ref{eq:PotMatrix1}) are 
sufficient to regularize the theory.

We implement a modified on-shell renormalization condition to fix the finite parts
of the counterterms. For example, the Feynman diagram from two insertions of 
$\left(\Phi^2_S-v^2\right)$ and the corresponding counterterm combine to a 
four-dimensional momentum space integral of the form
$$
\int \frac{d^4k}{(2\pi)^4}\,
\widetilde{v}_H(k)\widetilde{v}_H(-k)G_V(k^2)
$$
with
$$
G_V(k^2)=\int_0^1dx\,{\rm ln}\left[1-x(1-x)\frac{k^2}{M^2}\right] + c_V\,.
$$ 
Here $\widetilde{v}_H(k)$ is the Fourier transform of $\Phi^2_S-v^2$
and $c_V$ is the finite part of the counterterm coefficient. The
modified on-shell renormalization condition then adjusts $c_V$ such that
$G_V(M^2)=0$. That is, there is no quantum correction to the location
and residue of the pole due to the fluctuations. There are analogous
conditions for the counterterms $\int d^4x\,|D_\mu\Phi|^2$ and $\int
d^4x\, F_{\mu\nu}F^{\mu\nu}$. To simpify the calculation, we choose
conditions such that the mass of the gauge particle is unchanged,
meaning the masses remain equal at one-loop order. This choice comes
at the cost of introducing a correction to the residue of the pole.
An additional wavefunction renormalization to eliminate this
correction, as would be necessary to compare to a physical system,
would introduce a shift in the gauge mass.  However, the
effect of this shift only affects the VPE at higher loop orders.
Collecting  these terms, the finite counterterm
contribution to the VPE is
\begin{align}
\frac{E^{(0)}_{\rm CT}}{e^2v^2} &= \int_0^\infty \rho d\rho \left\{
\frac{n^2}{144}\left(\frac{16}{\pi}-3\sqrt{3}\right)\frac{g^{\prime2}}{\rho^2}
\right. \label{eq:Ect} \\ & \hspace{0.25cm} \left. 
+\left(\frac{1}{\pi}-2\sqrt{3}\right)
\left[h^{\prime2}+\frac{n^2}{\rho^2}h^2g^2 
+\frac{13}{8}\left(1-h^2\right)^2\right]
\right\}\, .
\nonumber
\end{align}

The Feynman diagram contribution arising from two insertions of $V_0$ is straightforward. 
We define Bessel-Fourier transforms
\begin{eqnarray}
I_A(k)&=&\int_0^\infty dr\, h(\rho) g(\rho) J_1(kr) \cr
I_H(k)&=&k\int_0^\infty rdr\,\left[1-h(\rho)\right]J_0(kr) \cr
\widetilde{v}(k)&=&\int_0^\infty
rdr\,\left[{h}^2(\rho)-1\right]J_0(kr) \,,
\end{eqnarray}
where $\rho=evr$, and obtain
\begin{eqnarray}
\frac{E_{V_0}}{e^2v^2}&=& \int_0^\infty \frac{k dk}{2\pi}
\left[n^2I^2_A(k)+I^2_H(k)+\frac{13e^2v^2}{8}\widetilde{v}^2(k)\right] \cr
&&\times \int_0^1dx\,{\rm ln}\left[1+x(1-x)\frac{k^2}{M^2}\right]\,.
\label{eq:EFDV2}
\end{eqnarray}
Finally, recall that we did not subtract the singular Born terms in Eq.~(\ref{eq:Escat})
but rather the fake boson analog. Hence we require the Feynman diagram energy
$$
E_{\rm fb}=\frac{c_B}{16\pi}\int_0^\infty kdk\,\widetilde{V}_f^2(k)
{\rm ln}\left[1+x(1-x)\frac{k^2}{M^2}\right]
$$
with $\widetilde{V}_f(k)=3\int_0^\infty rdr\,\left[\tanh^2(\kappa\rho)-1\right]J_0(kr)$,
corresponding to the fake boson potential.
Then $E^{(0)}_{\rm FD}=E_{V_0}+E_{\rm fb}$ completes our expression
for the VPE per  unit length of the vortex,
\begin{equation}
E_{\rm VPE}=E^{\rm scat.}_{\rm VPE}+E^{(0)}_{\rm FD}+E^{(0)}_{\rm CT}\,.
\label{eq:finalVPE}\end{equation}

\section{Numerical results for the VPE}

The numerical treatment is hampered by slow convergence due to the
logarithmic behavior of the subtracted Jost solution around the
center of the vortex and of $\left[\nu(t)\right]_V$ 
at small $t$. The detailed solutions to these problems will be
presented elsewhere~\cite{Graham:2021abc}.  One must also use a sum up
to a large number of partial waves $L$ together with an asymptotic
extrapolation to obtain accurate results for the sum over partial
waves~\cite{Graham:2019fzo}.

\begin{table}[ht]
\centerline{
\begin{tabular}{c||c|c|c|c}
& $n=1$ & $n=2$ & $n=3$ & $n=4$ \cr
\hline
$E^{(0)}_{\rm CT}$        & 0.0370 & 0.0786 & 0.1214 & 0.1651 \cr
$E^{(0)}_{\rm FD}$        & 0.0424 & 0.0315 & 0.0317 & 0.0335 \cr
\hline
                          & 0.0795 & 0.1101 & 0.1531 & 0.1986 \cr
$E^{\rm scat.}_{\rm VPE}$ &-0.0510 &-0.1939 &-0.3563 &-0.5251 \cr
\hline
$E_{\rm VPE}$             & 0.0284 &-0.0837 &-0.2033 &-0.3257 
\end{tabular}}
\caption{\label{tab:result}Various contributions to and the total VPE for 
different winding numbers~$n$. The fake boson potential is $V_f=3[\tanh^2(\kappa\rho)-1]$ 
with $\kappa=1,0.9,0.8,0.7$ for $n=1,2,3,4$, respectively. The third line contains the 
sum of the corresponding entries of the first two lines. All data are in units of $ev$.}
\end{table}

The results for the VPE and the various contributions that make it up
are displayed in Table~\ref{tab:result}. In units of $(ev)^2$ this
energy per unit length decreases by about 0.13 per winding number $n$,
with the $n=1$ VPE only slightly less than zero. A two-parameter fit
yields $E_{\rm VPE}\approx0.031-0.118(n-1)e^2v^2$
and thus $E_{\rm VPE}(n)-nE_{\rm VPE}(1)\approx 0.118(1-n)$.
Since the coefficient of the winding number is negative, the quantum
corrections stabilize the BPS-ANO vortex with higher winding number
and thus turn the system into a type I superconductor.

Because the biggest contribution comes from the (subtracted) scattering part,
a calculation based only on the leading Feynman diagrams would not be adequate.
Nevertheless, we also observe that the finite contribution due to our
modified on-shell renormalization, Eq.~(\ref{eq:Ect}), is significant
for $n>1$.

\section{Conclusions}

We have computed the one-loop quantum corrections to the energy per unit 
length of ANO vortices in scalar electrodynamics with spontaneous
symmetry breaking in the BPS case, where the masses of the scalar and gauge fields are 
equal. These corrections arise from the polarization of the spectrum of quantum fluctuations
in the classical vortex background. This vacuum polarization energy (VPE) is small because the 
small coupling approximation applies to electrodynamics 
with $e^2=4\pi/137\approx0.09$, but it becomes decisive for 
observables that vanish classically, such as
the binding energies of vortices with higher winding numbers in the
BPS case.

After clarifying a number of technical and numerical subtleties, we found that the dominant 
contribution to VPE of vortices stems from the full one-loop contribution, which cannot be 
computed from the lowest-order Feynman diagrams. On top of an infinite sum of Feynman diagrams
this contribution contains truely non-perturbative effects such as bound state energies that
are encoded within the exact Jost function.
Our numerical simulations for vortices with winding number up to four suggest that the quantum
energy weakly binds higher winding number BPS vortices. We have also seen that the VPE for the unit 
winding number vortex is very small, so that at first glance it appears to be compatible with zero
in the range of numerical errors. The potentially most important source for such errors
is the small radius behavior in channels that contain zero angular momentum components.
However, our error analysis suggests that any improvement of the data is likely to push that
VPE further away from zero by a few percent of the total VPE for all $n$ \cite{Graham:2021abc}.

To our knowledge this is the first study of a static soliton VPE
in a renormalizable model in four spacetime dimensions that
allows for a comparison of nontrivial winding numbers.  
Standard examples in one space dimension \cite{Ra82} inclue the kink
soliton, which has only a single nontrivial winding number, the
sine-Gordon soliton, which has classically degenerate solutions bound
by breather fluctuations, and the $\phi^6$ model soliton, which is 
destabilized by  quantum corrections \cite{Weigel:2019rhr}. 
The Skyrme model \cite{Skyrme:1961vq} in three space dimensions indeed
has static solitons  with different winding numbers for which one can
estimate quantum corrections~\cite{Scholtz:1993jg},
but unfortunately that model is not renormalizable.  It would be
interesting but more technically challenging to extend this
calculation to the case of a full 't Hooft-Polyakov monopole in three
dimensions.

\acknowledgments
N.\@ G.\@ is supported in part by the National Science Foundation (NSF)
through grant PHY-1820700.
H.\@ W.\@ is supported in part by the National Research Foundation of
South Africa (NRF) by grant~109497.

\bibliographystyle{apsrev}

\end{document}